\documentstyle{aipproc}
\begin{document}
\newcommand{\LI}{\hbox to\hsize}
\newcommand{\LLI}[1]{\LI{#1\hss}} \newcommand{\RLI}[1]{\LI{\hss#1}}
\newcommand{\CLI}[1]{\LI{\hss#1\hss}}
\def\null{\hbox{}}
%
%
%
\newcommand{\BEQ}{\begin{equation}}
\newcommand{\EEQ}{\end{equation}}
\newcommand{\norm}[1]{\mbox{$\left| \left| #1 \right| \right|$}}
\newcommand{\abs}[1]{\left| #1 \right|} 
\newcommand{\bra}[1]{\langle \left. #1 \right|}
\newcommand{\ket}[1]{\left| \right. #1 \rangle}
\newcommand{\bracket}[2]{\langle #1 \left| \right. #2 \rangle }
\newcommand{\DEG}[1]{\mbox{$ #1^{\rm o}$}}
\newcommand{\lappr}{\mbox{$\stackrel{<}{\sim}$}} 
\newcommand{\gappr}{\mbox{$\stackrel{>}{\sim}$}} 
\newcommand{\mr}[1]{\mbox{\rm #1}} 

%
\newcommand{\incircle}[1]{\mbox{{\hbox{$\bigcirc$}\kern-0.7em
\lower0.05ex\hbox{\mbox{{\scriptsize\rm #1}}}}}}
%
\newcommand{\eq}[1]{eq.~(\ref{#1})}
%
%
\newcommand{\ETC}{\mbox{\em etc.\/ }}
\newcommand{\VIZ}{\mbox{\em viz.\/ }}
\newcommand{\CF}{\mbox{\em cf.\/ }}
\newcommand{\IE}{\mbox{\em i.e. \/}}
\newcommand{\ETAL}{\mbox{\em et. al.\/ }}
\newcommand{\EG}{\mbox{\em e.g.\/ }}
\newcommand{\REG}{\mbox{\hskip.04ex \raisebox{0.5em}{\incircle{R}}\hskip.3em
}}
\newcommand{\mB}{\mbox{$ \mu_{B}$}}
%
\title{Neutrinos in Random Magnetic Fields:\\
The Problem of Measuring Magnetic Moments.}
\author{G. Domokos and S. Kovesi-Domokos}
\address{The Henry A. Rowland Department of Physics and Astronomy\\
The Johns Hopkins University\\
Baltimore, MD 21218.\thanks{E-mail: SKD@HAAR.PHA.JHU.EDU}}
\maketitle
\begin{abstract}
The existence of magnetic moments of neutrinos points to
physics beyond the standard model.  Given current upper limits,
terrestrial measurements are difficult or completely unfeasible. 
However, estimates of transition moments can be obtained from
observation of objects such as active galactic nuclei (AGN) by
means of neutrino telescopes. We describe the way of estimating the 
magnitudes  of transition moments from such observations. 
\end{abstract}
\section{Introduction and Summary}
The importance of measuring magnetic moments of neutrinos stems form the 
fact that the existence of such moments points to physics beyond the
standard model. To be sure, a minimal enlargement of the standard model
by means of right handed neutrinos alone leads to the existence of
magnetic moments induced by loops of charged gauge bosons and
charged leptons, \CF \cite{leeschrock}. However, due to the fact that
these moments are induced by means of higher order electroweak processes,
their magnitudes are very small. Typically, a diagonal moment is of
the order of magnitude,
\[ \mu_{\nu} \approx 3 \times 10^{-19}\frac{m_{\nu}/1eV}{\mu_{B}},\]
where $ \mu_{B}=e/(2m_{e})$ is the Bohr magneton. Transition moments also
contain mixing angles in their expressions, depending on the mixing schemes 
assumed. The important lesson, however, is that the standard model leads to
extremely small moments, beyond measurability for any experiment of the 
foreseeable future. Current experimental upper limits are much larger,
typically $\mu \leq 10^{-10} \mB$,\   \CF \cite{PDG}. Limits on transition
moments are, in general, model dependent and currently they are not
listed by the Particle Data Group. However, the consensus is that they are,
typically, larger by an order of magnitude or so.

Should magnetic moments in the range of the current upper limits be measured,
we would have an important low energy signal of the existence of  
physics beyond the
standard model, {\em with a very small background coming from the standard 
model itself.} Let us give a crude estimate of the relevant energy scale.
For purposes of illustration, we ignore flavor structure and mixing angles:
one hopes that this gives rise to errors of at most an order of magnitude.

The existence of an anomalous magnetic moment gives rise to a Pauli
term in
the low energy effective Lagrangian, \VIZ
\BEQ 
{\cal L}_{P} = \frac{e}{2m_{\nu}} \kappa \sigma_{\mu \nu} F^{\mu \nu},
\EEQ
where $m_{\nu}$ is the mass of the neutrino. The quantity $\kappa$ is
the ``low energy'' ($q_{\mu}\rightarrow 0$) limit of a spin flip
amplitude. Within factors of order unity, its magnitude is given by
an expression,
\BEQ
\kappa \simeq \frac{m_{\nu}}{\Lambda},
\EEQ
where $\Lambda$ is the characteristic energy scale of the process 
giving rise to the Pauli moment. The factor $m_{\nu}$ in the numerator
is present because an anomalous moment is generated by a spin flip
process. By putting in numbers, we discover for instance that
a magnetic moment of the order of $10^{-10}$\mB \   corresponds to an
energy scale of the order of $10^{4}$TeV.

Even though estimates of this type appear to be extremely naive ones, they work
quite well in cases where we know the mechanism by means of which a neutral
particle acquires a Pauli moment. Take the neutron as an example. Its
Pauli moment is: $\mu_{n}\approx 1.9 e/(2m_{n})$, \CF \cite{PDG}.
Using the previous estimates, we get that the characteristic mass scale is
\[ \Lambda_{n} \simeq 495 {\rm MeV}.\]
This value is quite close to $\Lambda_{QCD}$, which is expected to
characterize structure formation (\IE the formation of hadrons out of quarks) 
within QCD. 

The difficulty with small values of neutrino moments as given by
the upper limits quoted is that their measurement in a terrestrial experiment
is difficult or impossible. To illustrate this point, assume that
$\mu_{\nu} \simeq 10^{-10}$\mB.\   One can measure a magnetic moment by
inducing a spin flip in a magnetic field. Assuming the field to be a
homogeneous one, the distance over which a spin flip occurs on the average
is roughly $d\approx 1/\mu B$. A typical terrestrial magnet can 
maintain a field of the order of magnitude, $ B\simeq 1$Tesla.
Thus, the distance characterizing the spin flip of a moment of
$10^{-10}\mB$\   is $d\simeq 3\times 10^{4}$km. Clearly, no magnet
can be constructed on Earth which is that long.

The situation is better with transition moments. Depending on the nature of
neutrinos (Dirac or Majorana), a spin flip either leads to an 
active$\leftrightarrow$sterile conversion or to a flavor flip,
respectively. From the physical point of view, flavor flips
associated with spin flips are easier to observe and there are many
plausible models available which suggest that Nature may prefer
Majorana neutrinos over Dirac ones. For this reason, we shall
concentrate on Majorana neutrinos. It is to be noted that the observation
of a transition moment contains at least as much physical information as
the measurement of a diagonal moment does: hence there is no
disadvantage in looking for transition moments.

Keeping such arguments in mind, we suggested a way of measuring
transition moments of neutrinos in terrestrial experiments
utilizing facilities to be completed in the near future, namely,
long baseline oscillation experiments, \CF \cite{primakoff}.
Better sensitivities can be achieved by utilizing astrophysical
objects, such as an AGN, albeit at the cost of having to live with
greater uncertainties
of the properties of the source. 

Briefly, the idea is the following. Around an AGN, charged hadrons,
mostly pions are produced which subsequently decay, predominantly
into muons and $\nu_{\mu}$-s. The muons, being charged and long lived,
are expected to undergo a random walk in the surrounding plasma and
their decay products are unlikely to carry a substantial amount
of directional information about the source. By contrast, a $\nu_{\mu}$,
if left alone, would escape and reach a neutrino telescope, thus
carrying information about the source. In the presence of magnetic fields
in the emerging jets, however, flavor conversion can take place and this may
change the situation substantially. 

Presumably, charged particles in a jet are in a turbulent motion, hence, the
magnetic fields generated by them are chaotic. For the sake of simplicity,
we assume that the magnetic field present is a static, isotropic 
Gaussian random
field of zero mean and spatial correlation length $L$. This appears
to be a reasonable assumption, as long as the characteristic time scale of the
motion of the charged particles is longer than the time of passage of
neutrinos through the jet. Besides its correlation length, the Gaussian
field is characterized by the m.s. field, $\langle {\bf B}^{2}\rangle $. 

One expects that a neutrino traveling in a random magnetic field executes a
``random walk'' between its possible spin orientations. Hence, after a
while, the spin orientations become random; for a Majorana neutrino,
spin equilibrium also means flavor equilibrium. If the neutrino is produced
with a definite spin, its polarization is expected to be damped as a
function of distance, with a characteristic distance ${\cal D}$. 

We now argue that, within factors of order unity, the quantity ${\cal{D}}$
is uniquely determined: there is only one combination of the physical
quantities involved which is of the correct dimension.

Due to the fact that the coupling between a magnetic moment and a magnetic
field is proportional to ${\bf \mu \cdot B}$, the only combination in which
the field and the magnetic moment can enter is proportional to
$\mu^{2}\langle {\bf B}^{2}\rangle $, which is of  dimension  $\mbox{\rm length}^{-2}$.
Thus, the inverse spin flip length in  such a Gaussian field must be given
by an expression proportional to
\BEQ  
{\cal D}^{-1} =  \mu^{2} \langle {\bf B}^{2} \rangle L. 
\label{charlength}
\EEQ 

The important point to bear in mind is that the length given by \eq{charlength}
is rather short compared to typical jet sizes: hence, one expects the 
equilibrium to be established. In fact, by inserting r.m.s. fields
of the order of a few G, magnetic moments of the order of $10^{-10}$  to
$10^{-8}$\mB \   and correlation lengths of the order of a pc 
($\simeq  3\times 10^{13}$km), one ends up with ${\cal{D}} \ll L$, in
fact, of the order of merely a few times $10^{4}$km.

In the remainder of this talk, the basic ideas are illustrated on a simple, 
solvable model, along the lines described in ref.~\cite{pletrandom}.
The general theory, assuming an arbitrary number of flavors (and, in principle,
allowing arbitrary spins too), has been developed elsewhere \cite{festschrift}.
\section{A Tale of two Flavors}
As explained in the preceding section, we now consider the behavior 
of a single spin-1/2 field,
without paying detailed attention to the flavor structure.

In order to describe the average behavior of a neutrino in a random
magnetic field, one has to solve the dynamical equations governing
the propagation in an arbitrary magnetic field. The solution then has
to be averaged over the ensemble of  magnetic fields. 

We use the front form of
dynamics~\cite{dirac}. This
formulation of dynamics is advantageous in a situation in which one
considers the propagation of high energy particles ($E\gg m$, where
$m$ is the rest mass) and in which certain discrete symmetries, such as
$C$ and $P$ play no significant role. Clearly, the propagation of high
energy neutrinos falls into this category.

We begin with the usual Dirac Lagrangian of a particle in an external 
electromagnetic field, $F_{\mu \nu}$:
\BEQ \label{diraclagrangian}
L = \overline{\psi}\left(i \gamma^{\mu}\partial_{\mu} + m 
+ \frac{1}{2}\mu F^{\mu \nu}\sigma_{\mu \nu}\right)\psi
\EEQ
We work in the rest frame of the magnetic field.  Assuming the 
field to be a static one, we can set $F_{0i}=0$, $F_{ij}=\epsilon_{ijk}B_{k}$.
In the case of 
interest one has to solve the Dirac equation in an {\em arbitrary}
static magnetic field, since we want to average the solution over an
ensemble of  the $B_{i}$. No explicit solution is 
known for such a problem. However, we proceed to show that in the
{\em high energy limit} the problem can be solved in a closed form.

We introduce a coordinate system in which two of the coordinates
are null directions corresponding to characteristic lines of a relativistic 
wave equation, \VIZ:
\BEQ
t=\frac{1}{\sqrt{2}}\left( x^{0} - x^{3}\right), z=\frac{1}{\sqrt{2}}\left(
x^{0} + x^{3}\right) \quad {\rm and} \quad x^{A}; \quad (A=1,2).
\label{nullcoordinates}
\EEQ
Correspondingly, the metric is of the form,
\BEQ
g_{zt}=g_{tz}=1, \quad g_{AB}= -\delta_{AB},
\label{metric}
\EEQ
and all other components vanish.

A Dirac spinor can be decomposed along the null directions 
given in \eq{nullcoordinates} by introducing
the mutually orthogonal projectors,
\BEQ P_{t}=\frac{1}{2}\gamma_{t}\gamma^{t}, \quad P_{z}=\gamma_{z}\gamma^{z}
\label{nullprojectors}
\EEQ
In what follows, we use the shorthand,
\BEQ
\phi = P_{t}\psi, \quad \chi = P_{z}\psi
\label{shortnames}
\EEQ

It is a straightforward matter to decompose \eq{diraclagrangian}
according to the conjugate null directions and express it in terms of
the variables $\phi$ and $\chi$. The purpose of such an exercise is a very 
simple one. If, for the sake of definiteness, $t$ is regarded the ''time''
variable describing the dynamics of the system, only $\phi$ obeys
an equation containing $\partial_{t}$. Hence, the component of the 
Dirac spinor
corresponding to the conjugate null direction
 obeys only an equation of constraint. The constraint 
can be, in turn, solved before one attempts to attack the problem of dynamics.

After carrying out the decomposition of \eq{diraclagrangian} according
to the null directions, one finds: 
\begin{eqnarray}
L &=& \sqrt{2}\left[ \phi^{\dag}\left( i \partial_{t} -i \sqrt{2} \mu
\epsilon^{AB}\gamma_{A}B_{B}\right) \phi \right. \nonumber \\
  &+& \left. \chi^{\dag}\left( i\partial_{z} - i\sqrt{2} \mu
\epsilon^{AB} \gamma_{A}B_{B}\right) \chi \right] \nonumber \\
  &+& \frac{1}{\sqrt{2}}\left[ \phi^{\dag}\gamma^{z}\left(
i\gamma^{A}\partial_{A} +m - \frac{i}{\sqrt{2}}\mu B_{3}\,\epsilon_{AB}
\gamma ^{A} \gamma^{B}\right)\chi \right. \nonumber \\
  &+& \left.  \chi^{\dag}\gamma^{t}\left( i \gamma^{A}\partial_{A} + m 
 - \frac{i}{\sqrt{2}}\mu B_{3}\,\epsilon_{AB}
\gamma^{A}\gamma^{B}\right)\phi \right]
\label{decomposedlagrangian}
\end{eqnarray}

Variation of \eq{decomposedlagrangian} with respect to 
$\chi^{\dag}$ gives the constraint. The constraint can be solved in a
straight forward fashion and eliminated from the Lagrangian. The
result is conveniently written in Hamiltonian form:
\begin{eqnarray}
L & = & \pi \partial_{t}\phi -H \nonumber \\ 
H & = & -2\mu \phi^{\dag}\sigma^{A}B_{A}\phi \nonumber \\
  & + & \phi^{\dag}\left( -i \sigma_{B}\epsilon^{BC}p_{C} +
m -\mu \sqrt{2}B_{3}\sigma_{3}\right)\nonumber \\ 
& \times & \Omega  \left( -B^{A}\right)\nonumber \\
 & \times & \left( i\sigma_{R}\epsilon^{RS}p_{S}
 + m -\mu \sqrt{2} B_{3}\sigma{3}\right)\phi\nonumber \\
  &   &
\label{Hamiltonian}
\end{eqnarray}
Solving the constraint eliminates two components of
the original, four component Dirac spinor. Therefore, instead of the
original Dirac matrices one can use $2\times 2$ Pauli matrices.
One easily verifies that $-i\epsilon^{AB}\gamma_{B}\rightarrow
\sigma^{A}$ gives the correct representation. We also introduced the
Hermitean operator, $p_{A} = -i \partial_{A}$ for the transverse
degrees of freedom.

The canonical momentum is given by $\pi = i \sqrt{2}\, \phi^{\dag}$.
(Of course, the odd looking factor of $\sqrt{2}$ in the 
definition of the canonical momentum can be eliminated by
rescaling the time variable.)
In equation \eq{Hamiltonian}, $\Omega$ is an operator with matrix 
elements:
\BEQ
\bra{ z} \Omega \left( B^{A}\right) \ket{ z'} 
 = \frac{i}{\sqrt{2}}\exp\left(\mu \sqrt{2} 
\int_{z'}^{z} dz'\epsilon_{AB}\gamma^{A}B^{B}\right)\frac{1}{2}
\epsilon\left( z - z'\right)
\label{Omega}
\EEQ
All symbols of integration over $z$ have been
omitted. Eq.~\ref{Hamiltonian} is local in
$t$ and $x^{A}$; those arguments have been suppressed.

The Hamiltonian appearing in \eq{Hamiltonian} is  {\em exact\/.}
However, it is given by a rather complicated, non local  and
non linear expression:
this is the cost we have to pay for explicitly eliminating the 
constraint. We now argue that one can introduce physically reasonable
simplifications, as a result of which the problem becomes a manageable
one. First of all, we notice that the exponential appearing in \eq{Omega}
is of modulus one.  One expects that at large values 
of $|z - z'|$ the exponent
oscillates rapidly and thus contributes little to the Hamiltonian.
The dominant contribution is thus coming from small values of
 the difference of longitudinal coordinates. Hence, it is reasonable
to approximate the exponential in \eq{Omega} by 1.
In the remaining expression, one term is local in all variables and
the remaining ones are
proportional to $\epsilon\left( z- z'\right)$. In a Fourier
representation, \VIZ upon writing
\BEQ \phi \left(t, z, x^{A}\right) = \int dk \varphi\left( t, k, x^{A}
\right) \exp \left( -i kz\right)\EEQ
and
\BEQ \epsilon (z) = \frac{{\cal P}}{2\pi i}\int \frac{dk}{k} \exp (-i
kz), 
\EEQ
one recognizes that the non local terms in the Hamiltonian are proportional 
to negative 
powers of the longitudinal momentum, $k$.
(In the last equation
${\cal P}$ stands for the principal value.)
Hence,  at high energies ($k\gg m$) 
the Hamiltonian can be approximated by the local term.

Neglecting terms of $O\left( k^{-1}\right)$, the equation of motion for
the density matrix in coordinate representation  reads:
\begin{eqnarray}
-i \partial_{t}\bra{z,\vec{x}}\rho (t)\ket{z', \vec{x'}}& = &\mu \sqrt{2}
\vec{\sigma }\cdot\vec{B}\left(\vec{x},\frac{z-t}{\sqrt{2}}\right)\bra{z,\vec{x}}
\rho (t) \ket{z', \vec{x'}} \nonumber \\
 & - & \mu \sqrt{2}\bra{z ,\vec{x}}\rho
\left(t\right)\ket{z',\vec{x'}}\vec{\sigma}\vec{B}\left(\vec{x'},
\frac{z'-t}{\sqrt{2}}\right)
\label{eqofmotion}
\end{eqnarray}

In this equation, $\vec{x}$ stands for the transverse part of the
coordinate and $ \vec{\sigma}\cdot \vec{B}$ is the  two dimensional
scalar
product in transverse space. Of course,  the coordinate $x^{3}$ had to
be expressed by $ z$ and $t$; hence the $t$-dependence in the magnetic
field.

We choose the initial condition so as to describe a neutrino produced
at $\vec{x}=0$ and with a fixed value of $k$:
\BEQ
\bra{z, \vec{x}}\rho \left(0\right)\ket{z'\vec{x'}}
= \delta^{2}\left( \vec{x}\right)\delta^{2}\left( \vec{x'}\right)
\frac{\exp ik\left(z - z'\right)}{2\pi k}\rho_{s}\left(0\right),
\label{initialcondition} 
\EEQ
where $\rho_{s}(0)$ is the initial value of the spin density matrix.

The variable $k$ being large, the function
$\exp ik\left(z - z'\right)$ is  rapidly oscillating  unless
$z\approx z'$. Therefore, it is permissible to put $z=z'$ in the
coefficient  of the exponential in \eq{initialcondition}. Further, in
the
approximation used,  the
dynamics described by eq.~\eq{eqofmotion}
is independent of $k$ and of $\vec{x}$.
Therefore, the dependence of $\rho (t)$ on $k$ and $\vec{x}$ is
entirely
determined by the initial
condition. Thus, the dynamical equation reduces to an
equation 
involving the spin density matrix alone,  as in non relativistic
spin dynamics. Thus, from now on, we  omit the subscript $s$ and
we have:
\BEQ
-i\partial_{t}\rho \left( t \right) = \mu \sqrt{2}\left[ \vec{\sigma}\cdot
\vec{B}\left(\frac{z-t}{\sqrt{2}}\right), \rho \left( t\right)\right]
\label{spindynamics}
\EEQ
(Here and in what follows, $\vec{x}=0$ is understood.)

This equation can be solved by the standard time ordered series, \VIZ
\begin{eqnarray}
\rho\left(t\right)&=&\rho\left( 0\right)\nonumber \\
& +&i\mu \sqrt{2} \int_{0}^{t}dt'\left[\vec{\sigma}\cdot
\vec{B}\left(\frac{z-t'}{\sqrt{2}}\right)
,
\rho\left(0\right)
\right]\nonumber \\
 &+& \frac{\left( i\mu \sqrt{2}\right)^{2}}{2!}\int_{0}^{t}
dt'dt''T\left( \left[ \vec{\sigma}\cdot
\vec{B}\left(\frac{z-t'}{\sqrt{2}}\right),
\left[\vec{\sigma}\cdot \vec{B}\left(\frac{z-t''}{\sqrt{2}}\right),
\rho\left( 0 \right)\right]\right]\right)\nonumber \\ &+& \cdots
\label{timeordered}
\end{eqnarray}
We choose the initial condition as:
\BEQ \rho\left( 0\right) = \frac{1}{2}\left( 1 + S\sigma_{3}\right)
, \qquad \left(S^{2}\leq 1\right),
\EEQ
since neutrinos are produced with a definite helicity. (In  the case
of Dirac neutrinos, $S=\pm 1$, depending on whether a neutrino or
anti neutrino
is produced. In the case of Majorana neutrinos, $S$ may assume any
value between the limits stated above, depending on the production
mechanism.)

Next, we average the solution, \eq{timeordered} over the magnetic
field. We choose the generating functional of the moments as follows:
\begin{eqnarray}
Z[j] & = & \int {\cal D}B \exp -\left[ \frac{1}{2}\int
d^{3}xd^{3}x'B_{i}\left(x\right)C^{-1}_{ij}\left(x,x'\right)B_{j}
\left(x'\right)\right]\nonumber \\
& \times & \exp \int d^{3}x j_{i}
\left(x\right)B_{i}\left( x\right );\nonumber \\[1mm]
C_{ij}^{-1} & = & \frac{L}{4\pi \langle B^{2}\rangle}
\left(\delta_{ij}-\frac{\partial_{i}\partial_{j}}{\bigtriangledown^{2}}\right)
\left( L^{-2} - \bigtriangledown^{2}\right)^{2}\delta^{3}\left(x - x'\right).
\label{generatingfunctional}
\end{eqnarray} 
In the last equation, $L$ and $\langle B^{2}\rangle $ stand for the
correlation
length and mean square magnetic field, respectively. The measure is
normalized such that $Z\left[0\right]=1$.  
The transverse
projector is needed in order to make the correlation functions 
solenoidal. With the choice of the tensor $C^{-1}$ given in 
\eq{generatingfunctional}, the leading term in the long distance
behavior of the correlation function is $\propto \exp - \abs{x-x'}$.
In order to average equation \eq{timeordered} over the magnetic field,
one integrates over $B_{3}$ and sets the third component
of the source equal to zero. The  transverse generating functional
reads:
\begin{eqnarray}
Z_{T}&  = &\int {\cal D}\vec{B} \exp - \frac{L}{8\pi \langle B^{2}
\rangle } \int d^{3}x \left[ B^{A}(x) \left( \delta_{AB} - \frac{1}{2}
\frac{\partial_{A}\partial_{B}\left(x\right)}{\vec{\bigtriangledown}^{2}}\right)
B^{B}\right]\nonumber \\ 
& \times & \exp i \int d^{3} x \vec{j}\left(x\right)\cdot \vec{B}\left(x\right)
\label{transversegenerator}
\end{eqnarray}
  
We now notice that in  \eq{timeordered}, terms containing
odd
powers of $\mu$ are also odd in $B^{A}$. Therefore, in the limit
$\vec{j} \rightarrow 0$ the average of those terms vanishes. The even
terms in the series are obtained by taking the appropriate functional
derivatives of \eq{transversegenerator}. All of them are expressed in
terms of multiple time integrals of $C_{ij}\left(\abs{t-t'}\right)$
and its powers: those integrations are easily performed. It is
sufficient to illustrate the procedure for the second order term in 
\eq{timeordered}.

Carrying out the integrations, one gets:
\[
 - \mu^{2} \frac{1}{2} S\langle \int_{0}^{t} dt' dt'' \left[
\vec{\sigma}
\cdot \vec{B},\left[
\vec{\sigma}\cdot \vec{B}, \sigma_{3}\right]\right]\rangle
= - \mu^{2}\sigma_{3} \langle B^{2}\rangle tL \left( 1 - \exp
-\frac{t}{L}
\right)
\]
For large times the result in the last equation is
just proportional to $t$. The higher order terms follow a similar
pattern. The end result is:
\BEQ
\langle \rho \left( t \right) \rangle \sim
\frac{1}{2}\left( 1 + S\sigma_{3}\exp - \frac{t}{T}\right),
\label{asymptote}
\EEQ
with
\[ \frac{1}{T} = 2\mu^{2} \langle B^{2}\rangle L . \]

Thus we arrive at the remarkable result that in the random field
the behavior of the helicities is an {\em ergodic} one: irrespective
of what the initial density matrix was, for $t\gg T$, the helicities
are equally distributed. Furthermore, as conjectured, the characteristic
length over which the flavor equilibrium is established is about the
characteristic length ${\cal D}$, \CF \eq{charlength} in the
previous section.
\section{Discussion}
The result conjectured in the first section and verified  
within the framework of a simplified model
in the previous one indicates that if neutrino telescopes will detect
neutrinos originating from AGN (and perhaps from other astronomical objects),
it is likely that all flavors will occur with equal probability. 
By contrast, if the arriving neutrinos are predominantly $\nu_{\mu}$-s,
one concludes that neutrinos are Dirac particles and the  magnetic
fields present around the source cause mainly an 
active$\leftrightarrow$sterile conversion. Such a conversion decreases
the intensity of active neutrinos by about a factor of two. 
({\em In principle\/,} knowing the flux of
neutrinos produced at the source, one could verify the
existence of the conversion. However, current flux estimates are, 
by far, not  accurate enough for such a purpose.)

Even though we conjectured the existence of an equilibrium
of spin orientation on the basis of a simple intuitive argument and
verified the conjecture within the framework of a rather simple minded
model, we believe the result to be a rather general one. In fact,
there exists in the literature a rather substantial number of
papers in which conclusions similar to ours have been reached in
a variety of physical circumstances (the early Universe, the 
interior of the Sun, \ETC\/). For  an incomplete, but
representative sample, see \EG refs.~\cite{elmfors}, \cite{nicolaidis}, \cite{enquist}.
The probability distributions assumed for the magnetic fields
are quite different (\EG white noise with a finite spatial extension,
zero spatial correlation length with randomly oriented domains, \ETC).
All  the distributions, however, have the properties discussed
in Section I: the magnetic field has a vanishing mean, it is
characterized by $\langle B^{2}\rangle$ and by some length.
Hence, the dimensional argument given in Sec. I is
applicable; consequently, the characteristic distance found in
every work of this type agrees with ${\cal D}$ within factors of
order one. 
Consequently, it is reasonable to expect that our results are
rather robust ones and thus, testing them by means of
flavor sensitive neutrino telescopes is of substantial physical
interest.
\section*{Acknowledgement}
The work reported here has been carried out during the authors'
visit at the Dipartimento di Fisica, Universit\'{a} di Firenze.
We thank Roberto Casalbuoni, Director of the Department,
for the warm hospitality extended to us and to Bianca Monteleoni
for useful conversations on neutrino telescopes.

We also wish to thank Per Osland and all of our colleagues at the
Department of Physics, 
University of Bergen for organizing and running a very successful
and enjoyable meeting.

\end{document}